# A Compressed Sensing Wire-Tap Channel

Galen Reeves, Naveen Goela, Nebojsa Milosavljevic, and Michael Gastpar[†]
Department of Electrical Engineering and Computer Sciences
University of California, Berkeley

*Abstract*—A multiplicative Gaussian wire-tap channel inspired by compressed sensing is studied. Lower and upper bounds on the secrecy capacity are derived, and shown to be relatively tight in the large system limit for a large class of compressed sensing matrices. Surprisingly, it is shown that the secrecy capacity of this channel is nearly equal to the capacity without any secrecy constraint provided that the channel of the eavesdropper is strictly worse than the channel of the intended receiver. In other words, the eavesdropper can see almost everything and yet learn almost nothing. This behavior, which contrasts sharply with that of many commonly studied wiretap channels, is made possible by the fact that a small number of linear projections can make a crucial difference in the ability to estimate sparse vectors.

## I. INTRODUCTION

Following Shannon's theory in 1949 of information-theoretic secrecy [1], Wyner introduced the wiretap channel in 1975 [2]. In the wiretap setting, a sender Alice wishes to communicate a message to a receiver Bob over a main channel but her transmissions are intercepted by an eavesdropper Eve through a secondary wiretap channel. The present paper analyzes a multiplicative Gaussian wiretap channel inspired by compressed sensing. The input to the channel is a $p$-length binary vector. The channel output is a linear transform of the input after it has first been corrupted by multiplicative white Gaussian noise. We analyze the setting where Bob and Eve observe different linear transforms characterized by two different channel matrices.

Secrecy via compressed sensing schemes has received little attention from an information-theoretic viewpoint. In prior work, authors consider using a sensing matrix as a key (unknown to the eavesdropper) for both encryption and compression [3]. Privacy via compressed sensing and linear programming decoding was explored in [4]. By contrast, this paper assumes that the sensing matrices are known (non-secret); as a special case, Eve's sensing matrix might correspond to a subset of the rows of Bob's channel matrix. Our analysis shows that certain channel matrices, inspired by compressed sensing, allow for secrecy rates that are nearly equal to the main channel capacity even if Eve's capacity is large.

### A. Channel Model

Outlined in Fig. 1, the multiplicative Gaussian wiretap channel with binary vector input is characterized by

$$\mathbf{Y} = A_b W \mathbf{X}, \quad (1)$$
$$\mathbf{Z} = A_e W \mathbf{X}, \quad (2)$$

[†]Also with the School of Computer and Communication Sciences, EPFL, Lausanne, Switzerland.

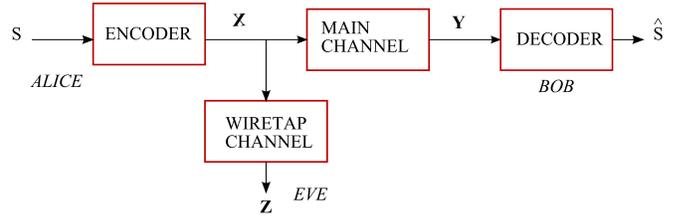

Fig. 1. *(Multiplicative Gaussian Wiretap Channel)* For each block length $n$, Alice transmits a sequence of $n$ binary valued support vectors $\mathbf{X} \in \{0,1\}^p$ over a main channel characterized by a matrix transform so that Bob receives $\mathbf{Y} = A_b W \mathbf{X}$. The eavesdropper Eve receives $\mathbf{Z} = A_e W \mathbf{X}$.

where $\mathbf{X} \in \{0,1\}^p$ is the transmitted signal, and $\mathbf{Y} \in \mathbb{R}^{m_b}$, $\mathbf{Z} \in \mathbb{R}^{m_e}$ are the received real-valued signals at the legitimate user and eavesdropper, respectively. A related channel model in [5] also involves a wire-tap setting with binary input and real-valued output. The dimensions of the channel satisfy $0 \leq m_e < m_b < p/2$. The linear mixing parameters, matrices $A_b \in \mathbb{R}^{m_b \times p}$ and $A_e \in \mathbb{R}^{m_e \times p}$, are fixed and known to all parties. The randomness of the channel is derived from $W \in \mathbb{R}^{p \times p}$, a diagonal matrix whose values are i.i.d. Gaussian random variables with mean zero and variance one. The channel is assumed to be memoryless between channel uses.

### B. Secrecy Capacity

Alice selects a message $S_n \in [1 : 2^{npR}]$, where $R$ represents a normalized rate, and wishes to communicate reliably with Bob while keeping the message secret from Eve. A $(2^{npR}, n)$ secrecy code for the multiplicative wiretap channel consists of the following: (1) A message set $[1 : 2^{npR}]$; (2) A randomized encoder that generates a codeword $\mathbf{X}^n(S_n)$, $S_n \in [1 : 2^{npR}]$, according to $P_{\mathbf{X}^n|S_n}$; (3) A decoder that assigns a message $\hat{S}_n(\mathbf{Y}^n)$ to each received sequence $\mathbf{Y}^n \in \mathcal{Y}^n$. The message $S_n$ is a random variable with entropy satisfying

$$\lim_{n \to \infty} \frac{H(S_n)}{np} = R. \quad (3)$$

A secrecy code is reliable if

$$\lim_{n \to \infty} \Pr[\hat{S}_n(\mathbf{Y}^n) \neq S_n] = 0. \quad (4)$$

A secrecy code is secret if the information leakage rate tends to zero as block length $n \to \infty$,

$$\lim_{n \to \infty} \frac{I(\mathbf{Z}^n; S_n)}{n} = 0. \quad (5)$$

Note that this leakage rate is not normalized by $p$. A normalized rate $R$ is achievable if there exists a sequence of

$(2^{npR}, n)$ secrecy codes satisfying both Eqn. (4) and Eqn. (5). The secrecy capacity $C_s$ is the supremum over all achievable rates.

## C. Outline

To analyze the secrecy capacity $C_s$, we first develop bounds as a function of the channel matrices $A_b$ and $A_e$. We then analyze these bounds for certain random matrices in the large system limit where $m_b/p \to \rho_b$ and $m_e/p \to \rho_e$ as $p \to \infty$ for fixed constants $0 \le \rho_e \le \rho_b \le 1/2$. Lower bounds on the secrecy capacity, corresponding to Wyner's coding strategy for discrete memoryless channels are developed in Section II-A. Corresponding upper bounds are derived in Section II-B. Section II-C provides an improved upper bound under a certain encoding constraint on Alice. Proofs are given in Section III.

## D. Notations

For a matrix $A \in \mathbb{R}^{m \times p}$ and vector $\mathbf{x} \in \{0,1\}^p$, we use $A(\mathbf{x})$ to denote the matrix formed by concatenating the columns indexed by $\mathbf{x}$, and we use $A(i)$ to denote the $i$th column of $A$. Also, we use $\mathcal{X}_k^p$ to denote the set of all binary vectors $\mathbf{x} \in \{0,1\}^p$ with exactly $k$ ones. We use $H_2(x) = -x\log x - (1-x)\log(1-x)$ to denote binary the binary entropy function. We use $\log$ to denote the logarithm with base two and $\ln$ to denote the logarithm with the natural base.

## II. MAIN RESULTS

Csiszar and Korner showed in [6] that the secrecy capacity of a discrete memoryless wiretap channel is given by

$$C_s = \max_{(U,\mathbf{X})} \left[\tfrac{1}{p} I(U;\mathbf{Y}) - \tfrac{1}{p} I(U;\mathbf{Z})\right] \quad (6)$$

where the auxiliary random variable $U$ satisfies the Markov chain relationship: $U \to \mathbf{X} \to (\mathbf{Y}, \mathbf{Z})$. It can be verified that this is also the secrecy capacity when the channels have discrete inputs and continuous outputs (see e.g. [5]).

In some special cases, the secrecy capacity can be computed easily from (6). For example, if $A_b$ and $A_e$ correspond to the first $m_b$ and $m_e$ rows of the $p \times p$ identity matrix respectively, then it is straightforward to show that

$$C_s = \begin{cases} \frac{m_b}{p} - \frac{m_e}{p}, & \text{if } m_e < m_b \\ 0 & \text{if } m_e \ge m_b \end{cases}. \quad (7)$$

In this case, the secrecy capacity happens to be the difference of the individual channel capacities; thus as $m_e$ approaches $m_b$ the secrecy capacity tends to zero. In the following sections we will develop bounds for a class of matrices inspired by compressed sensing. Interestingly, we will see that the secrecy behavior of these matrices differs greatly from the behavior shown in (7).

## A. Lower Bounds

We say that a matrix $A \in \mathbb{R}^{m \times p}$ is *fully linearly independent* (FLI) if the span of each submatrix $\{A(\mathbf{x}) \in \mathbb{R}^{m \times m-1} : \mathbf{x} \in \mathcal{X}_{m-1}^p\}$ defines a unique linear subspace of $\mathbb{R}^m$. Examples of FLI matrices include the first $m$ rows of the $p \times p$ discrete cosine transform matrix or, with probability one, any matrix whose entries are drawn i.i.d. from a continuous distribution. A counter example is given by the first $m$ rows of the $p \times p$ identity matrix.

Our first result, which is proved in Section III-A, gives a general lower bound on the secrecy capacity for any FLI matrices.

**Theorem 1.** *Suppose that $A_b$ and $A_e$ are fully linearly independent. If $m_e < m_b$, then the secrecy capacity is lower bounded by*

$$\begin{aligned} C_s \ge{} & \tfrac{1}{p}\log\binom{p}{m_b-1} - \tfrac{1}{2p}\log\det(\tfrac{1}{p}A_e A_e^T) \\ & + \sum_{\mathbf{x} \in \mathcal{X}_{m_b-1}^p} \tfrac{1}{\binom{p}{m_b-1} 2p} \log\det\left(\tfrac{1}{m_b-1}A_e(\mathbf{x})A_e(\mathbf{x})^T\right). \end{aligned} \quad (8)$$

The lower bound in Theorem 1 is derived by evaluating the right hand side of (6) when $\mathbf{X}$ is distributed uniformly over the set $\mathcal{X}_{m_b-1}^p$. We note that the condition $m_e < m_b$ is necessary to obtain a nontrivial lower bound since the secrecy capacity may be equal to zero otherwise.

Unfortunately, the bound in Theorem 1 is difficult to compute if $m_b$ and $p$ are large. One way to address this issue is to analyze the behavior for a random matrix (random matrices are denoted via boldface, uppercase letters). The following result is proved in Section III-B.

**Theorem 2.** *Suppose that $A_b$ is fully linearly independent and $\mathbf{A}_e$ is a random matrix whose elements are i.i.d. $\mathcal{N}(0,1)$. If $m_e < m_b$ then the expectation of the secrecy capacity is lower bounded by*

$$\begin{aligned} \mathbb{E}_{\mathbf{A}_e}[C_s] \ge{} & \tfrac{1}{p}\log\binom{p}{m_b-1} - \tfrac{m_e}{2p}\log\left(\tfrac{m_b-1}{p}\right) \\ & - \tfrac{\log e}{2p} \sum_{i=1}^{m_e} \left[\psi\left(\tfrac{p-i+1}{2}\right) - \psi\left(\tfrac{m_b-i}{2}\right)\right] \end{aligned} \quad (9)$$

*where $\psi(x) = \Gamma'(x)/\Gamma(x)$ is Euler's digamma function.*

One benefit of Theorem 2 is that the bound is independent of the realization of the matrix $\mathbf{A}_e$ and can be analyzed directly. An illustration of the bound is shown in Figure 2 as a function of $m_e/p$ for various values of $p$ with $m_b/p$ held fixed. Remarkably, as $p$ becomes large, the lower bound in Theorem 2 remains bounded away from zero for all values of $m_e$ strictly less than $m_b$. This behavior is in stark contrast to the secrecy capacity shown in (7).

One shortcoming of Theorem 2, is that the bound holds only in expectation, and it is possible that it is violated for a constant fraction of matrices $\mathbf{A}_e$. The next result, which is proved in Section III-C, shows that, in the asymptotic setting, the limit of the bound (9) holds for almost every realization of $\mathbf{A}_e$. We use the notation $\{A^{(p)} \in \mathbb{R}^{m^{(p)} \times p}\}$ to denote a sequence of matrices indexed by the number of columns $p$.

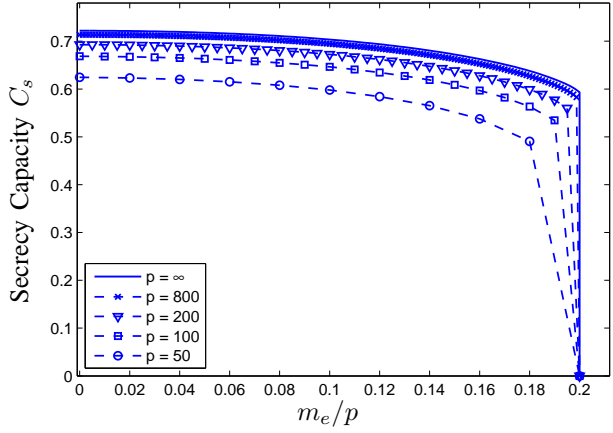

Fig. 2. Illustration of the lower bound in Theorem 2 on the expected secrecy capacity $\mathbb{E}_{\mathbf{A}_e}[C_s]$ as a function of $m_e$ for various values of $p$ when $A_b$ is fully linearly independent, $\mathbf{A}_e$ is a random matrix whose elements are i.i.d. $\mathcal{N}(0,1)$, and $m_b/p = 0.2$.

**Theorem 3.** *Suppose that $\{A_b^{(p)} \in \mathbb{R}^{m_b^{(p)} \times p}\}$ is a sequence of linearly independent matrices and $\{\mathbf{A}_e^{(p)} \in \mathbb{R}^{m_e^{(p)} \times p}\}$ is a sequence of random matrices whose elements are i.i.d. $\mathcal{N}(0,1)$. If $m_e^{(p)} > m_b^{(p)}$ and $m_b^{(p)}/p \to \rho_b$ and $m_e^{(p)}/p \to \rho_e$ as $p \to \infty$ where $0 \le \rho_e \le \rho_b \le 1/2$, then the asymptotic secrecy capacity is lower bounded by*

$$\liminf_{p\to\infty} C_s \ge H_2(\rho_b) - \tfrac{1}{2}\Big[(1-\rho_e)\log\big(\tfrac{1}{1-\rho_e}\big) \\ - (\rho_b - \rho_e)\log\big(\tfrac{\rho_b}{\rho_b-\rho_e}\big)\Big] \quad (10)$$

*almost surely.*

Theorem 3 provides a concise characterization of the lower bound in the asymptotic setting. The bound is illustrated in Figure 2 in the case $p = \infty$. Since the secrecy capacity can be equal to zero if $m_e^{(p)} = m_b^{(p)}$, Theorem 3 shows that there is a *discontinuity* in the asymptotic secrecy capacity as a function of $\rho_e$.

### B. Upper Bounds via Channel Capacity

This section considers the capacity of Bob's channel which is denoted $C_b$. We note that this capacity gives us an upper bound on the secrecy capacity.

Upper bounding the capacity is more technically challenging than lower bounding the secrecy capacity, since the optimal distribution on $\mathbf{X}$ may depend nontrivially on channel matrix $A_b$. The following result, which is proved in Section III-D, serves as a starting point.

**Theorem 4.** *If $A_b$ is fully linearly independent, then the channel capacity of Bob's channel is upper bounded by*

$$C_b \le \tfrac{1}{p}\max\big(\log\binom{p}{m_b-1}, \max_{m_b \le k \le p} \tilde{c}(k)\big) + \tfrac{\log p}{p} \quad (11)$$

*where*

$$\tilde{c}(k) = \max_{1 \le i \le p} \tfrac{m_b}{2}\log\big(\tfrac{1}{m_b}\|A_b(i)\|^2\big) \\ - \max_{\mathbf{x}\in\mathcal{X}_k^p} \tfrac{1}{2}\log\det(\tfrac{1}{k}A_b(\mathbf{x})A_b(\mathbf{x})^T). \quad (12)$$

Although it is tempting to consider the expectation of (11) with respect to a random matrix (as we did for Theorem 2), this is difficult since the maximization in (12) occurs inside the expectation.

Our next result, which is proved in Section III-E, leverages the strong concentration properties of the Gaussian distribution to characterize the asymptotic capacity for Gaussian matrices.

**Theorem 5.** *Suppose that $\{\mathbf{A}_b^{(p)} \in \mathbb{R}^{m_b^{(p)} \times p}\}$ is a sequence of random matrices whose elements are i.i.d. $\mathcal{N}(0,1)$. If $m_b^{(p)}/p \to \rho_b$ where $0 < \rho_b \le 1/2$, then the asymptotic channel capacity of Bob's channel is given by*

$$\lim_{p\to\infty} C_b = H_2(\rho_b) \quad (13)$$

*almost surely.*

Theorem 5 shows that the strategy used in our lower bounds, namely choosing $\mathbf{X}$ uniformly over $\mathcal{X}_{m_b-1}^p$ achieves the capacity of Bob's channel in the asymptotic setting. What is remarkable is that for this same input distribution, Eve learns very little about what is being sent, even if her channel matrix is equal to the first $m_e$ rows of Bob's channel matrix.

### C. Improved Upper Bound for a Restricted Setting

We say that the distribution is *symmetric* if $\Pr[\mathbf{X} = \mathbf{x}] = \Pr[\mathbf{X} = \tilde{\mathbf{x}}]$ for all $\mathbf{x}, \tilde{\mathbf{x}}$ such that $\sum_{i=1}^p x_i = \sum_{i=1}^p \tilde{x}_i$. The following result is proved in Section III-F.

**Theorem 6.** *Suppose that $\{\mathbf{A}_b^{(p)} \in \mathbb{R}^{m_b^{(p)} \times p}\}$ and $\{\mathbf{A}_e^{(p)} \in \mathbb{R}^{m_e^{(p)} \times p}\}$ are sequences of random matrices whose elements are i.i.d. $\mathcal{N}(0,1)$. If $m_b^{(p)} > m_e^{(p)}$ and $m_b^{(p)}/p \to \rho_b$, and $m_e^{(p)}/p \to \rho_e$ where $0 \le \rho_e \le \rho_b \le 1/2$, and if Alice is restricted to use coding strategies that induce a symmetric distribution on $\mathbf{X}$, then the asymptotic secrecy capacity is upper bounded by*

$$\limsup_{p\to\infty} C_s \le H(X | WX + \sqrt{\rho_b/\rho_e}V) \quad (14)$$

*almost surely where $X \sim \text{Bernoulli}(\rho_b)$, $W \sim \mathcal{N}(0,1)$ and $V \sim \mathcal{N}(0,1)$ are independent random variables.*

The bound in Theorem 6 is strictly less than the channel capacity $H_2(\rho_b)$ for all $\rho_e > 0$, and can be computed easily using numerical integration. We suspect that this result also holds without the symmetry restriction on $\mathbf{X}$.

### D. Illustration of Bounds

The bounds on the asymptotic secrecy capacity given in Theorems 3, 5, are 6 and illustrated in Fig. 3 as a function of the size parameter $\rho_e$ of the eavesdropper channel. The bounds correspond to the setting where the elements of the matrices are i.i.d. Gaussian. Note that the lower bound on the secrecy

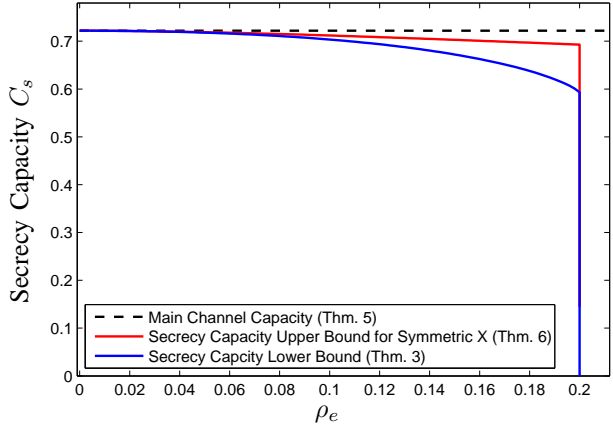

Fig. 3. Bounds on the asymptotic (normalized) secrecy capacity $C_s$ of the multiplicative Gaussian wiretap channel as a function of $\rho_e$ when $\rho_b = 0.2$ and $\mathbf{A}_b$ and $\mathbf{A}_e$ are random matrices whose elements are i.i.d. $\mathcal{N}(0,1)$.

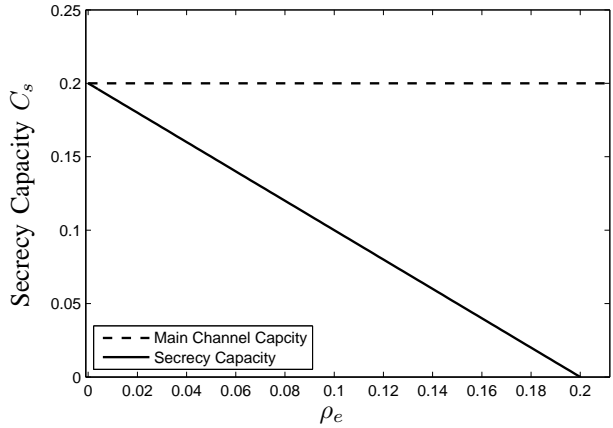

Fig. 4. The (normalized) secrecy capacity $C_s$ of the multiplicative Gaussian wiretap channel as a function of $\rho_e$ when $\rho_b = 0.2$ and $A_b$ and $A_e$ correspond to the first rows of the identity matrix.

capacity is nearly equal to that of the main channel for all $\rho_e < \rho_b$.

For comparison, the secrecy capacity for the special case where $A_b$ and $A_e$ correspond to the first rows of the $p \times p$ identity matrix are shown in Fig. 4. In this case, the secrecy capacity is equal to the difference of the main channel capacity and eavesdropper capacity.

## III. PROOFS

### A. Proof of Theorem 1

Let $U = \mathbf{X}$ where $\mathbf{X}$ is distributed uniformly over $\mathcal{X}_{m_b-1}^p$. Since $A_b$ is fully linearly independent, the probability that $\mathbf{Y}$ is in the range space of $A_b(\tilde{\mathbf{x}})$ for any $\tilde{\mathbf{x}} \in \mathcal{X}_{m_b-1}^p$ not equal to the true vector $\mathbf{X}$ is equal to zero. Thus, $H(\mathbf{X}|\mathbf{Y}) = 0$ and

$$I(U;\mathbf{Y}) = I(\mathbf{X};\mathbf{Y}) = H(\mathbf{X}) = \log\binom{p}{m_b-1}. \quad (15)$$

Next, since $A_e$ is fully linearly independent and the number of nonzero values in $\mathbf{X}$ is strictly greater than the rank of $A_e$, it can be verified that both $\mathbf{Z}$ and $\mathbf{Z}|\mathbf{X}$ have probability densities. (Note that the condition $m_b > m_e$ is critical here, since $\mathbf{Z}$ does not have a density otherwise.) Thus we can write

$$I(U;\mathbf{Z}) = I(\mathbf{X};\mathbf{Z}) = h(\mathbf{Z}) - h(\mathbf{Z}|\mathbf{X}) \quad (16)$$

where $h(\cdot)$ denotes differential entropy (see e.g. [7]). The entropy $h(\mathbf{Z})$ can be upper bounded as

$$h(\mathbf{Z}) \leq \max_{\tilde{\mathbf{Z}}\,:\,\mathbb{E}[\tilde{\mathbf{Z}}\tilde{\mathbf{Z}}] = \mathbb{E}[\mathbf{Z}\mathbf{Z}^T]} h(\tilde{\mathbf{Z}})$$
$$\leq \tfrac{1}{2}\log\left((2\pi e)^{m_e} \det(\mathbb{E}[\mathbf{Z}\mathbf{Z}^T])\right) \quad (17)$$
$$= \tfrac{1}{2}\log\left((2\pi e\,(\tfrac{m_b-1}{p}))^{m_e} \det(A_e A_e^T)\right) \quad (18)$$

where (17) follows from the fact that the Gaussian distribution maximizes differential entropy and (18) follows from the fact that

$$\mathbb{E}[\mathbf{Z}\mathbf{Z}^T] = \tfrac{m_b-1}{p} A_e A_e^T. \quad (19)$$

The conditional entropy $h(\mathbf{Z}|\mathbf{X})$ is given by

$$h(\mathbf{Z}|\mathbf{X}) = \mathbb{E}\left[\tfrac{1}{2}\log\left((2\pi e)^{m_e} \det(A_e(\mathbf{X}) A_e(\mathbf{X})^T)\right)\right] \quad (20)$$

where we used the fact that, conditioned on any realization $\mathbf{X} = \mathbf{x}$, $\mathbf{Z}$ is a (non-degenerate) Gaussian random vector with covariance matrix $A_e(\mathbf{x}) A_e(\mathbf{x})^T$. Combining (15), (16), (18), and (20) with the expression of the secrecy capacity given in (6) completes the proof of Theorem 1.

### B. Proof of Theorem 2

It is straightforward to show that $\mathbf{A}_e$ is fully linearly independent with probability one. Since the secrecy rate is bounded, it thus follows from Theorem 1 and the linearity of expectation that

$$\mathbb{E}[C_s] \geq \tfrac{1}{p}\log\binom{p}{m_b-1} - \tfrac{m_e}{2p}\log\left(\tfrac{m_b-1}{p}\right)$$
$$- \tfrac{1}{2p}\mathbb{E}\left[\log\det(\mathbf{A}_e \mathbf{A}_e^T)\right]$$
$$+ \sum_{\mathbf{x}\in\mathcal{X}_{m_b-1}^p} \tfrac{1}{\binom{p}{m_b-1} 2p}\mathbb{E}\left[\log\det\left(\mathbf{A}_e(\mathbf{x}) \mathbf{A}_e(\mathbf{x})^T\right)\right]. \quad (21)$$

Using well known properties of random Gaussian matrices (see e.g. [8, pp. 99-103]) shows that

$$\mathbb{E}\left[\log\det(\mathbf{A}_e \mathbf{A}_e^T)\right] = m_e + \log e \sum_{i=1}^{m_e} \psi(\tfrac{p-i+1}{2})$$

$$\mathbb{E}\left[\log\det(\mathbf{A}_e(\mathbf{x}) \mathbf{A}_e(\mathbf{x})^T)\right] = m_e + \log e \sum_{i=1}^{m_e} \psi(\tfrac{m_b-i}{2})$$

where the second equality holds for every $\mathbf{x} \in \mathcal{X}_{m_b-1}^p$. Plugging these expressions into (21) completes the proof.

## C. Proof of Theorem 3

Since $\mathbf{A}_e$ is fully linearly independent with probability one, it is sufficient to consider the asymptotic behavior of the bound in Theorem 1. If $\mathbf{X}$ is a random vector distributed uniformly over $\mathcal{X}^p_{m_b-1}$ then $\mathbf{A}_e \mathbf{A}_e^T$ and $\mathbf{A}_e(\mathbf{X})\mathbf{A}_e(\mathbf{X})^T$ are $m_e \times m_e$ Wishart matrices with $p$ and $m_b - 1$ degrees of freedom respectively. Using Lemma 6 in the Appendix gives

$$\lim_{p \to \infty} \tfrac{1}{p} \log \det(\tfrac{1}{p}\mathbf{A}_e \mathbf{A}_e^T) = \mu(\rho_e)$$
$$\lim_{p \to \infty} \tfrac{1}{p} \log \det(\tfrac{1}{m_b-1}\mathbf{A}_e(\mathbf{X})\mathbf{A}_e^T(\mathbf{X})) = \rho_b \, \mu(\rho_e/\rho_b)$$

almost surely where $\mu(r) = (1-r)\ln\left(\tfrac{1}{1-r}\right) - r \log e$. Thus,

$$\lim_{p \to \infty} \left[ \tfrac{m_e}{p} \log\left(\tfrac{m_b-1}{p}\right) \tfrac{1}{p} \log \det(\mathbf{A}_e \mathbf{A}_e^T) \right.$$
$$\left. - \sum_{\mathbf{x} \in \mathcal{X}^p_{m_b-1}} \tfrac{1}{\binom{p}{m_b-1}p} \log \det(\mathbf{A}_e(\mathbf{x})\mathbf{A}_e(\mathbf{x})^T) \right]$$
$$= \lim_{p \to \infty} \left[ \tfrac{m_e}{p} \log\left(\tfrac{m_b-1}{p}\right) \tfrac{1}{p} \log \det(\mathbf{A}_e \mathbf{A}_e^T) \right.$$
$$\left. - \tfrac{1}{p} \log \det(\mathbf{A}_e(\mathbf{X})\mathbf{A}_e(\mathbf{X})^T) \right] \quad (22)$$
$$= \mu(\rho_b) - \rho_b \, \mu(\rho_e/\rho_b)$$
$$= (1-\rho_e)\log\left(\tfrac{1}{1-\rho_e}\right) - (\rho_b - \rho_e)\log\left(\tfrac{\rho_b}{\rho_b - \rho_e}\right) \quad (23)$$

almost surely where the substitution in (22) is justified by the fact that the expectation of $\tfrac{1}{m_b} \log \det(\tfrac{1}{m_b}\mathbf{A}(\mathbf{X})\mathbf{A}_e(\mathbf{X})^T)$ with respect to both $\mathbf{A}$ and $\mathbf{X}$ is bounded uniformly for all $p$ (see the proof of Theorem 2).

Combining (23) with the well known fact that

$$\lim_{p \to \infty} \tfrac{1}{p} \log \binom{p}{m_b-1} = H_2(\rho_b) \quad (24)$$

completes the proof of Theorem 3.

## D. Proof of Theorem 4

Let $K = \sum_{i=1}^{p} X_i$ denote the number of ones in $\mathbf{X}$. Then,

$$I(\mathbf{X}; \mathbf{Y}) = I(\mathbf{X}; \mathbf{Y}|K) + I(\mathbf{Y}; K) \quad (25)$$
$$\leq I(\mathbf{X}; \mathbf{Y}|K) + \log p \quad (26)$$
$$\leq \max_{0 \leq k \leq p} I(\mathbf{X}; \mathbf{Y}|K=k) + \log p \quad (27)$$

where (25) follows from the chain rule for mutual information, (26) follows from the fact that $I(\mathbf{Y}; K) \leq H(k) \leq \log p$, and (27) follows from expanding the term $I(\mathbf{X}; \mathbf{Y}|K)$. If we define

$$c(k) = \max_{\mathbf{X} \in \mathcal{X}^p_k} I(\mathbf{X}; \mathbf{Y})$$

then we have

$$\max_{\mathbf{X}} I(\mathbf{X}; \mathbf{Y}) \leq \max_{0 \leq k \leq p} c(k) + \log p. \quad (28)$$

To complete the proof, we split the maximization over $k$ into two cases. For $0 \leq k < m_b$ we use the simple bound

$$\max_{0 \leq k < m_b} c(k) \leq \max_{\mathbf{X} \in \mathcal{X}_k \,:\, 0 \leq k < m_b} H(\mathbf{X}) = \log \binom{p}{m_b-1}.$$

For $m_b \leq k \leq p$ we use the following lemma.

**Lemma 1.** *If $m_b \leq k \leq p$, then $c(k) \leq \tilde{c}(k)$ where $\tilde{c}(k)$ is given in (12).*

*Proof:* Let $\mathbf{X}$ have any distribution on $\mathcal{X}^p_k$ where $m_b \leq k \leq p$. Since $A_e$ is fully linearly independent, both $\mathbf{Y}$ and $\mathbf{Y}|\mathbf{X}$ have probability densities and we can write

$$I(\mathbf{X}; \mathbf{Y}) = h(\mathbf{Y}) - h(\mathbf{Y}|\mathbf{X}) \quad (29)$$

where $h(\cdot)$ denotes differential entropy (see e.g. [7]). The entropy $h(\mathbf{Y})$ can be upper bounded as

$$h(\mathbf{Y}) \leq \max_{\tilde{\mathbf{Y}}\,:\,\mathbb{E}[\|\tilde{\mathbf{Y}}\|^2]=\mathbb{E}[\|\mathbf{Y}\|^2]} h(\tilde{\mathbf{Y}})$$
$$= \tfrac{m_b}{2}\log\left(2\pi e \tfrac{1}{m_b}\mathbb{E}[\|\mathbf{Y}\|^2]\right) \quad (30)$$
$$\leq \max_{1 \leq i \leq p} \tfrac{m_b}{2}\log\left(2\pi e \tfrac{k}{m_b}\|A_b(i)\|^2\right) \quad (31)$$

where (30) follows from the fact that an isotropic Gaussian vector maximizes differential entropy, and (31) follows from the fact that

$$\mathbb{E}[\|\mathbf{Y}\|^2] = \mathbb{E}\big[\mathbb{E}[\|\mathbf{Y}\|^2|\mathbf{X}]\big]$$
$$\leq \max_{\mathbf{x}} \mathbb{E}[\|\mathbf{Y}\|^2|\mathbf{X}=\mathbf{x}]$$
$$= \max_{\mathbf{x}} \operatorname{tr}\big(A_b(\mathbf{x})A_b(\mathbf{x})^T\big)$$
$$\leq \max_{1 \leq i \leq p} k\|A_b(i)\|^2.$$

The conditional entropy $h(\mathbf{Y}|\mathbf{X})$ is lower bounded by

$$h(\mathbf{Y}|\mathbf{X}) = \mathbb{E}\left[\tfrac{1}{2}\log\left((2\pi e)^{m_b} \det(A_b(\mathbf{X})A_b(\mathbf{X})^T)\right)\right]$$
$$\geq \min_{\mathbf{x} \in \mathcal{X}^p_k} \tfrac{1}{2}\log\left((2\pi e)^{m_b} \det(A_b(\mathbf{x})A_b(\mathbf{x})^T)\right) \quad (32)$$

where we used the fact that, conditioned on any realization $\mathbf{X} = \mathbf{x}$, $\mathbf{Y}$ is a (non-degenerate) Gaussian random vector with covariance matrix $A_b(\mathbf{x})A_b(\mathbf{x})^T$. Combining (29), (31) and (32) completes the proof of Theorem 4. ∎

## E. Proof of Theorem 5

Since $\mathbf{A}_e$ is fully linearly independent with probability one, it is sufficient to consider the asymptotic behavior of the bound in Theorem 4. The limit of the first term in the maximization is given by (24). To evaluate the second term, we use the following technical lemmas whose proofs are given in the Appendices A and B.

**Lemma 2.**

$$\limsup_{p \to \infty} \max_{1 \leq i \leq p} \frac{1}{m_b}\|\mathbf{A}_b(i)\|^2 \leq 1 \quad (33)$$

*almost surely.*

**Lemma 3.** *If $k \geq m_b$ and $k/p \to \kappa$ where $\rho_b \leq \kappa \leq 1$, then*

$$\liminf_{p \to \infty} \min_{\mathbf{x} \in \mathcal{X}^p_k} \tfrac{1}{p}\log \det(\tfrac{1}{k}\mathbf{A}_b(\mathbf{x})\mathbf{A}_b(\mathbf{x})^T) \geq \kappa\,\mu(\rho_b/\kappa) \quad (34)$$

*almost surely where $\mu(r) = (1-r)\log\left(\tfrac{1}{1-r}\right) - r\log e$.*

The convergence show in Lemmas 2 and 3 leads immediately to the following asymptotic upper bound on the term $\tilde{c}(k)$ defined in (12):

$$\limsup_{p\to\infty} \max_{m_b < k \le p} \frac{1}{p}\tilde{c}(k)$$
$$\le \max_{\rho_b \le \kappa \le 1} \frac{1}{2}\left[\rho_b \log e - (\kappa - \rho_b)\log\left(\frac{\kappa}{\kappa - \rho_b}\right)\right]$$
$$= \frac{1}{2}\rho_b \log e$$

almost surely. Since $H_2(\rho_b) > \frac{1}{2}\rho_b \log e$ for all $\rho_b \in (0, 1/2)$, we conclude that the asymptotic capacity is upper bounded by $H_2(\rho_b)$. The achievable strategy outlined in the proof of Theorem 1 shows that $H_2(\rho_b)$ is also achievable which concludes the proof of Theorem 5.

### F. Proof of Theorem 6

Let $K = \sum_{i=1}^p X_i$. Then, for any pair $(U, \mathbf{X})$ such that $U \to \mathbf{X} \to (\mathbf{Y}, \mathbf{Z})$, we have

$$I(U; \mathbf{Y}) - I(U; \mathbf{Z})$$
$$= I(\mathbf{X}; \mathbf{Y}) - I(\mathbf{X}; \mathbf{Z}) + I(\mathbf{X}; \mathbf{Z}|U) - I(\mathbf{X}; \mathbf{Y}|U) \quad (35)$$
$$\le I(\mathbf{X}; \mathbf{Y}|K) - I(\mathbf{X}; \mathbf{Z}|K)$$
$$\quad + I(\mathbf{X}; \mathbf{Z}|U, K) - I(\mathbf{X}; \mathbf{Y}|U, K) + 2\log p \quad (36)$$
$$\le \max_{0 \le k \le p} \Delta(k) + 2\log p \quad (37)$$

where

$$\Delta(k) = I(\mathbf{X}; \mathbf{Y}|K=k) - I(\mathbf{X}; \mathbf{Z}|K=k)$$
$$\quad + I(\mathbf{X}; \mathbf{Z}|U, K=k) - I(\mathbf{X}; \mathbf{Y}|U, K=k).$$

We now consider two cases. For the case $m_b \le k \le p$, we use the upper bound

$$\Delta(k) \le I(\mathbf{X}; \mathbf{Y}|K=k) + I(\mathbf{X}; \mathbf{Z}|U, K=k)$$
$$\le I(\mathbf{X}; \mathbf{Y}|K=k) + I(\mathbf{X}; \mathbf{Z}|K=k)$$

which follows from the non-negativity of mutual information and the data processing inequality. Following the steps outlined in the proofs of Theorems 4 and 5 shows that

$$\limsup_{p\to\infty} \max_{\mathbf{X} \in \mathcal{X}_k^p : m_b \le k \le p} \frac{1}{p}\Delta(k) \le \frac{1}{2}(\rho_b + \rho_e)\log e \le \rho_b \log e$$
(38)

almost surely. (Note that this step does not require the symmetry assumption.)

Alternatively, for the case $0 \le k < m_b$, we use the bound

$$\Delta(k) \le H(\mathbf{X}|\mathbf{Z}, K=k) + H(\mathbf{X}|\mathbf{Y}, K=k)$$

which follows from the non-negativity of entropy and the fact that conditioning cannot increase entropy. Since $\mathbf{A}_b$ is fully linearly independent almost surely, it follows from the proof of Theorem 1 that $H(\mathbf{X}|\mathbf{Y}, K=k)$ is equal to zero almost surely. To characterize the asymptotic behavior of the remaining term, $H(\mathbf{X}|\mathbf{Z}, K=k)$, we use the following lemma which is proved in Appendix C.

**Lemma 4.** *Suppose that $\mathbf{X}$ is symmetric. If $0 \le k < m_b$ and $k/p \to \kappa$ where $0 \le \kappa \le \rho_b$, then*

$$\limsup_{p\to\infty} \frac{1}{p} H(\mathbf{X}|\mathbf{Z}, K=k) \le g(\kappa, \rho_e) \quad (39)$$

*almost surely where*

$$g(\kappa, \rho_e) = H(X|WX + \sqrt{\kappa/\rho_e}V) \quad (40)$$

*and $X \sim \text{Bernoulli}(\kappa)$, $W \sim \mathcal{N}(0, 1)$ and $V \sim \mathcal{N}(0, 1)$ are independent random variables.*

Noting that $g(\kappa, \rho_e)$ is nondecreasing in $\kappa$, we obtain the asymptotic upper bound

$$\limsup_{p\to\infty} C_s \le \max\big(g(\rho_b, \rho_e), \rho_b\big). \quad (41)$$

It can be verified numerically that this maximum occurs at $g(\rho_b, \rho_e)$ for all $\rho_b \in (0, 1/2)$ which completes the proof of Theorem 6.

## IV. CONCLUSION

Bounds on the secrecy capacity $C_s$ for the multiplicative Gaussian wiretap channel were analyzed for a class of channel matrices inspired by compressed sensing. One natural extension of the observations in this paper would be to consider the combined effects of additive noise in vector wiretap channels.

## APPENDIX

### A. Proof of Lemma 2

Note that the column magnitudes $\|\mathbf{A}_b(i)\|^2, i = 1, 2, \cdots, p$ are i.i.d. chi-square random variables with $m_b$ degrees of freedom. Thus for any $\epsilon \in (0, 1/2)$, the chi-square concentration inequality in Lemma 5 in Appendix D gives

$$\Pr[\max_{1 \le i \le p} \frac{1}{m_b}\|\mathbf{A}_b(i)\|^2 \ge 1 + \epsilon] \le p\exp(-\frac{3}{16}\lceil \rho_b p \rceil \epsilon^2) \quad (42)$$

which decays exponentially rapidly with $p$ as $p \to \infty$.

### B. Proof of Lemma 3

For each $\mathbf{x} \in \mathcal{X}_k^p$, let

$$N(\mathbf{A}_b, \mathbf{x}) = \frac{1}{k}\log\det\left(\frac{1}{k}\mathbf{A}(\mathbf{x})\mathbf{A}(\mathbf{x})^T\right). \quad (43)$$

By the union bound, and the symmetry of $\mathbf{A}_b$ we have

$$\Pr\left[\min_{\mathbf{x} \in \mathcal{X}_k^p} N(\mathbf{A}_b, \mathbf{x}) \le t\right] \le \binom{p}{k}\Pr[N(\mathbf{A}, \mathbf{x}) \le t], \quad (44)$$

for any arbitrary $\mathbf{x} \in \mathcal{X}_k^p$. Using the bound $\binom{p}{k} \le (pe/k)^k$ and Lemma 6 in Appendix D, shows that for any $\epsilon > 0$,

$$\limsup_{p\to\infty} \frac{1}{p\ln p}\ln\Pr[N(\mathbf{A}, \mathbf{x}) \le \mu(\rho_b/\kappa) - \epsilon] \le -\epsilon$$

which suffices to prove almost sure convergence.

## C. Proof of Lemma 4

Let $\hat{\mathbf{X}} = \frac{1}{m_e}\mathbf{A}_e^T\mathbf{Z}$. Then, we have

$$H(\mathbf{X}|\mathbf{Z}, K=k) \leq H(\mathbf{X}|\hat{\mathbf{X}}, K=k) \tag{45}$$

$$\leq \sum_{i=1}^{p} H(X_i|\hat{\mathbf{X}}, K=k) \tag{46}$$

$$\leq \sum_{i=1}^{p} H(X_i|\hat{X}_i, K=k) \tag{47}$$

where (45) follows from the data processing inequality, (46) follows from the chain rule and (47) follows from the fact that conditioning cannot increase entropy.

Next, we observe that $\hat{X}_i$ can be written as

$$\hat{X}_i = \frac{1}{m_e}\sum_{i=1}^{p}\langle\mathbf{A}_e(i), \mathbf{A}_e(j)\rangle W_j X_j \tag{48}$$

$$= \frac{\|\mathbf{A}_e(i)\|^2}{m_e}W_i X_i + \sigma_i(\mathbf{A}_e, \mathbf{X})V \tag{49}$$

where $V \sim \mathcal{N}(0,1)$ is independent of $W_i, X_i$ and $\sigma_i^2(\mathbf{A}_e, \mathbf{X})$ where

$$\sigma_i^2(\mathbf{A}_e, \mathbf{X}) = \frac{1}{m_e}\sum_{j\neq i}\langle\mathbf{A}_e(i), \mathbf{A}_e(j)\rangle^2 X_j. \tag{50}$$

Using standard chi-square inequality, it is straightforward to show that

$$\lim_{p\to\infty}\max_{1\leq i\leq p}\left|\frac{\|\mathbf{A}_e(i)\|^2}{m_e} - 1\right| = 0 \tag{51}$$

almost surely. With a bit more work, and the use of the fact that, by the symmetry constraint, $\mathbf{X}$ is distributed uniformly over $\mathcal{X}_k^p$ it can also be shown that

$$\lim_{p\to\infty}\max_{1\leq i\leq p}|\sigma_j^2(\mathbf{A}_e, \mathbf{X}) - \tfrac{\kappa}{\rho_e}| = 0 \tag{52}$$

almost surely. Thus, we conclude that the empirical distribution of the pairs $(X_i, \hat{X}_i)$ converges weakly almost surely to the distribution on $(X, WX + \sqrt{\kappa/\rho_e}V)$, which concludes the proof.

## D. Technical Lemmas

**Lemma 5** ( [10]). *If $X$ is a chi-square random variable with $n$ degrees of freedom then for all $\epsilon \in (0, 1/2)$,*

$$\Pr[X \geq d(1+\epsilon)] \leq \exp\left(-\tfrac{3}{16}d\epsilon^2\right). \tag{53}$$

**Lemma 6.** *Let $\mathbf{W}$ be an $m \times m$ Wishart random matrix with $n \geq m$ degrees of freedom. If $m/n \to \rho \in (0,1]$ as $n \to \infty$, then for any $\epsilon > 0$,*

$$\limsup_{n\to\infty} \tfrac{1}{n\ln n}\ln\Pr\left[\left|\tfrac{1}{n}\log\det(\tfrac{1}{n}\mathbf{W}) - \mu(\rho)\right| > \epsilon\right] \leq -\tfrac{\epsilon}{\log e}$$

*where*

$$\mu(\rho) = \begin{cases} (1-\rho)\log\left(\tfrac{1}{1-\rho}\right) - \rho\log e, & \text{if } 0 < \rho < 1 \\ -\log e, & \text{if } \rho = 1 \end{cases}. \tag{54}$$

*Proof:* We begin with a one-sided bound. For any $r > 0$ we have

$$\Pr\left[\tfrac{1}{n}\ln\det\left(\tfrac{1}{n}\mathbf{W}\right) \leq t\right]$$
$$= \Pr\left[\ln\det(\mathbf{W}) \leq nt + m\log n\right]$$
$$= \Pr\left[\left(\det(\mathbf{W})\right)^{-r} \geq \exp(-rnt + m\log n)\right]$$
$$\leq \exp(rnt + rm\log n)\mathbb{E}\left[\left(\det(\mathbf{W})\right)^{-r}\right] \tag{55}$$

where (55) follows from Markov's inequality. If $r < (n-m)/2$, then it can be shown (see e.g. [8, pp. 99-103]) that

$$\mathbb{E}\left[\left(\det(\mathbf{W})\right)^{-r}\right] = \exp(-M(r))$$

where

$$M(r) = rm\ln(2) + \sum_{i=0}^{m-1}\left[\ln\Gamma(\tfrac{n-i}{2}) - \ln\Gamma(\tfrac{n-i}{2} - r)\right].$$

If $r$ is an integer then we use the relation

$$\ln\Gamma(z) - \ln\Gamma(z - r) = \sum_{i=1}^{r}\ln(z - i)$$

to obtain

$$M(r) = rm\ln(2) + \sum_{i=0}^{m-1}\sum_{j=1}^{r}\ln\left(\tfrac{n-i}{2} - j\right)$$

$$= rm\ln n + \sum_{i=0}^{m-1}\sum_{j=1}^{r}\ln\left(\tfrac{n-i-2j}{n}\right).$$

Plugging this back into (55) gives

$$\ln\Pr\left[\tfrac{1}{n}\ln\det(\tfrac{1}{n}\mathbf{W}) \leq t\right] < rnt - \sum_{i=0}^{m-1}\sum_{j=1}^{r}\ln\left(\tfrac{n-i-2j}{n}\right).$$

We now consider what happens as $n \to \infty$. If $r = \ln n$ then it is straightforward to show that

$$\lim_{n\to\infty}\frac{1}{rn}\sum_{i=0}^{m-1}\sum_{j=1}^{r}\ln\left(\tfrac{n-i-2j}{n}\right) = \frac{\mu(\rho)}{\log e},$$

and thus

$$\limsup_{n\to\infty}\tfrac{1}{n\ln n}\ln\Pr\left[\tfrac{1}{n}\ln\det\left(\tfrac{1}{n}\mathbf{W}\right) \leq t\right] \leq t - \tfrac{\mu(\rho)}{\log e}.$$

To prove the other side of the bound, we use the same steps as before to obtain

$$\ln\Pr\left[\tfrac{1}{n}\ln\det(\tfrac{1}{n}\mathbf{W}) \geq t\right] < \sum_{i=0}^{m-1}\sum_{j=1}^{r}\log\left(\tfrac{n-i+2j}{n}\right) - rnt.$$

Letting $r = \ln n$ leads to

$$\limsup_{n\to\infty}\tfrac{1}{n\ln n}\ln\Pr\left[\tfrac{1}{n}\ln\det\left(\tfrac{1}{n}\mathbf{W}\right) \geq t\right] \leq \tfrac{\mu(\rho)}{\log e}, -t.$$

Changing the base of the logarithms concludes the proof of Lemma 6. ∎

ACKNOWLEDGMENT

This work was supported in part by ARO MURI No. W911NF-06-1-0076, and in part by 3TU.CeDICT: Centre for Dependable ICT Systems, The Netherlands.